\documentclass{PoS}
\usepackage{graphicx}
\usepackage{caption}
\usepackage{pinlabel}
\title{Charm production in association with an electroweak gauge boson at the LHC }

\ShortTitle{Charm production in association with an electroweak gauge boson at the LHC }

\author{\speaker{Eleni Vryonidou}\thanks{In collaboration with James Stirling.}\\
        Cambridge University (UK)\\
        E-mail: \email{hv222@hep.phy.cam.ac.uk}}

\abstract{The production of charm quark jets in association with electroweak gauge bosons at the LHC can be used as a tool to constrain quark 
parton distribution functions. Motivated by recent measurements at the Tevatron and LHC, we calculate cross sections for  
$W/Z+c$, comparing these to  $W/Z+\textrm{jet}$, for various PDF sets. The cross-section differences can be understood in terms 
of the different underlying PDFs, with the strange quark distribution being particularly important for $W+c$ production. 
We discuss appropriately defined ratios and comment on how the measurements at the 
LHC can be used to extract information on the strange and charm content of the proton at high $Q^2$ scales. }

\FullConference{XXI International Workshop on Deep-Inelastic Scattering and Related Subject -DIS2013,\\
		22-26 April 2013\\
		Marseilles,France}

\begin{document}
\section{Introduction}
 Electroweak measurements at the LHC can be used to extract information on the strange and charm content of the proton, particularly at high $Q^2$ scales. A first determination of the strange quark distribution from LHC data has been performed by ATLAS in \cite{Aad:2012sb}, where information is extracted by studying the total $W^{\pm},Z$ cross sections. 
In addition to the total $W^{\pm},Z$ cross sections, a more direct way of extracting information on the strange quark distribution is by studying the associated production of $W$ bosons and charm quark jets \cite{Stirling:2012vh}. Charm tagging in production of electroweak gauge bosons at hadron colliders can provide important information on strange 
and charm quark PDFs, complementary to that obtained by tagging charm quarks in the final state in deep 
inelastic scattering experiments~\cite{Lai:2007dq}. In particular, CDF and D0   
have measured the cross section for charm quarks produced in association with $W$ bosons \cite{Aaltonen:2007dm,Abazov:2008qz,Aaltonen:2012wn}, with an accuracy of $\sim$30\%. The LHC can provide a more precise measurement, and indeed the 
CMS collaboration has recently performed a similar study \cite{CMS:2011lqa,CMS:qwa}, 
using long-lifetime tagging to identify the charm jets. ATLAS has very recently released the first results for the same process in \cite{ATLAS-CONF-2013-045}.

\begin{figure}[h]
\centering
\includegraphics[scale=0.5]{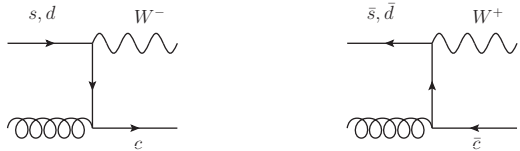} 
\caption{Feynman diagrams for $W^\pm+c(\bar c)$ production at LO.}
\label{Feync}
\end{figure}

At leading order (LO), the $t-$channel Feynman diagrams for $W+c$ production are shown in Fig.~\ref{Feync}. 
The dominant contribution comes from strange quark -- gluon scattering, as the corresponding 
down-quark contribution is strongly Cabibbo suppressed.

Charm production in association with $Z$ bosons can be used to extract information on the charm quark PDF. For both 
$W+c$ and $Z+c$ production at hadron colliders, the strange and charm quark PDFs are probed at much higher 
factorisation scales $Q^2$  ($\sim 10^4$~GeV$^2$) than in the traditional 
determinations from DIS, i.e. $\nu s \to \mu^- c(\to \mu^+)$ and $e c \to e c (\to \mu^+)$ where typically 
$Q^2 \sim 10^{0-2}$~GeV$^2$. Taken together, the measurements therefore also test DGLAP evolution for these quark flavours.

We study $W +c$-jet production in the context of the CMS analysis~\cite{CMS:2011lqa}, 
analysing the different quark contributions and comparing the predictions of various widely-used PDF sets. 
We also discuss $Z+c$-jet production, which should be measurable 
at the LHC.

\section{$W+c$ at the LHC }
In order to facilitate the comparison with the CMS analysis we use two cross-section ratios introduced by CMS~\cite{CMS:2011lqa}:
\begin{equation}
R_c^{\pm}=\frac{\sigma(W^++\bar{c})}{\sigma(W^-+c)} \,\,\,\, \textrm{and} \,\,\,\, 
 R_c=\frac{\sigma(W+c)}{\sigma(W+\textrm{jet})}.
\label{eq:Wcratios}
\end{equation}
The advantage of using ratios is that many of the theoretical and experimental uncertainties cancel. 
Note that the charm charge asymmetry ratio $R_c^\pm \equiv 1$ at the Tevatron. We calculate the cross sections at NLO pQCD
using MCFM~\cite{MCFM}, applying the CMS cuts \cite{CMS:2011lqa} to the final-state: 
$p_T^{\rm{jet}}>20$ GeV, $|\eta^{\rm{jet}}|<2.1$, $p_T^{\rm{lepton}}>25$ GeV, $|\eta^{\rm{lepton}}|<2.1$, 
$R^{jj}=0.5$, $R^{lj}=0.3$, where $R^{jj}$ and $R^{lj}$ are respectively the jet-jet and lepton-jet minimum separation parameters. 
Five different NLO PDF sets are used: CT10~\cite{Lai:2010vv}, MSTW2008~\cite{Martin:2009iq}, 
NNPDF2.1~\cite{Ball:2011uy}, GJR08~\cite{Gluck:2007ck} and ABKM09~\cite{Alekhin:2009ni}, as implemented in LHAPDF~\cite{LHAPDF}.
The renormalisation and factorisation scales are set to $M_W$, i.e. $\mu_R = \mu_F = M_W$,
although the cross-section ratios are rather 
insensitive to this choice. 

The results are summarised in Table \ref{fractions2} where we also include:
\begin{equation}
R^{\pm}=\frac{\sigma(W^++ \textrm{jet})}{\sigma(W^-+\textrm{jet})} .
\label{eq:Wpmrat}
\end{equation} 
\begin{table}[h]
\renewcommand*{\arraystretch}{1.4}
\begin{center}
    \begin{tabular}{ | c | c | c | c |}
    \hline 
   Ratio  & $R_c^{\pm}$ & $R_c$  & $R^{\pm}$   \\ \hline
   CT10  & $0.953^{+0.009}_{-0.007}$ & $0.124^{+0.021}_{-0.012}$ & $1.39^{+0.03}_{+0.03}$ \\ 
  MSTW2008NLO &  $0.921^{+0.022}_{-0.033}$  & $0.116^{+0.002}_{-0.002}$ & $1.34^{+0.01}_{-0.01}$ \\ 
NNPDF2.1NLO & 0.944$\pm$0.008  & 0.104$\pm$0.005  & 1.39$\pm$0.02\\
GJR08 & 0.933$\pm$0.003 & 0.099$\pm$0.002 & 1.37$\pm$0.02\\
ABKM09 & 0.933$\pm$0.002 & 0.116$\pm$0.003 & 1.39$\pm$0.01\\ \hline
   \end{tabular}
\end{center}
\caption{Comparison of results at NLO using different PDF sets, including 68\%cl (asymmetric, where available) PDF errors.}
\label{fractions2}
\end{table}

For reference, we note the values of $R_c^{\pm}$ and $R_c$ measured by CMS \cite{CMS:2011lqa}:
\begin{eqnarray}
R_c^{\pm}=0.92\pm 0.19\,(\textrm{stat.})\pm0.04\,(\textrm{syst.})\\
R_c=0.143\pm0.015\,(\textrm{stat.})\pm 0.024\,(\textrm{syst.})
\end{eqnarray}
CMS has updated the $W+c$ analysis with the 2011 data set in \cite{CMS:qwa}. The analysis involves a different set of selection cuts. The result they obtained is $R_c^{\pm}=0.954\pm0.025\,(\textrm{stat.})\pm0.001\,(\textrm{syst.})$ for $p_T^{\rm{lepton}}>25$ GeV. Both systematic and statistical errors have significantly decreased.

If only the strange quark contributed to $W+c$ production, then any deviation of $R_c^\pm$ from $1$ 
would imply an asymmetry between $s$ and $\bar s$. However even if $s = \bar s$, the fact that $\bar d<d$ will automatically give 
$R_c^\pm<1$ through the Cabibbo suppressed $d-$quark contribution. Schematically, at LO we expect
\begin{equation}
R_c^{\pm}\sim \frac{\bar{s}+|V_{dc}|^2\bar{d}}{s+|V_{dc}|^2d},
\label{eq:sd}
\end{equation}
with $V_{dc}$=0.225. This leads to a suppression by a factor of 20 of the $d-$quark contribution to the cross section. 

\begin{figure}[h]
 \begin{minipage}[b]{0.5\linewidth}
\centering
\includegraphics[trim=1.2cm 0 0 0,scale=0.45]{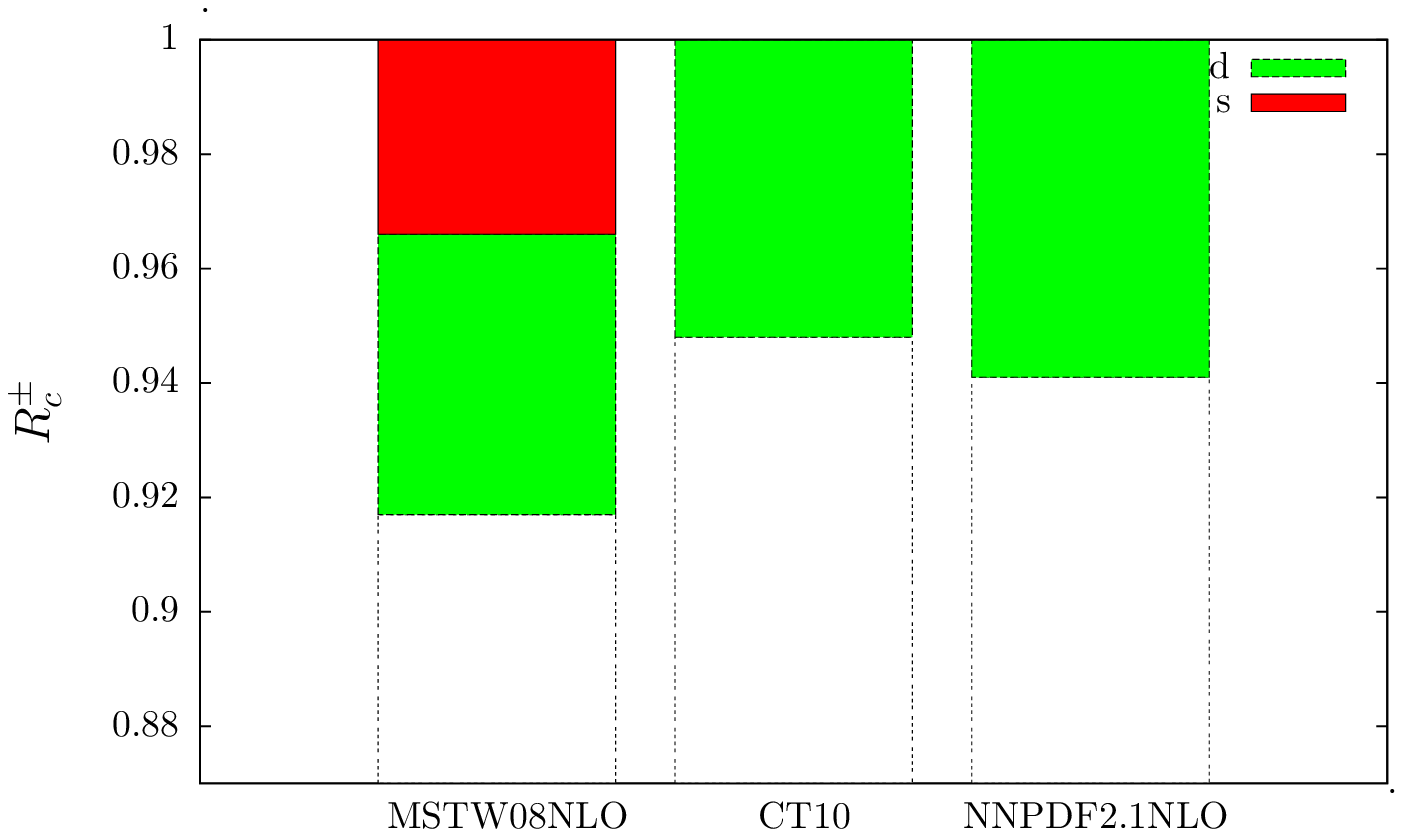}
\caption{Effect of initial-state $s$ and $d$ quarks on $R_c^{\pm}$ using NLO PDFs (LO processes).}
\label{NLOpdf1}
\end{minipage}
\hspace{0.5cm}
\begin{minipage}[b]{0.5\linewidth}
 \centering
 \includegraphics[trim=2.3cm 0 0 0,scale=0.45]{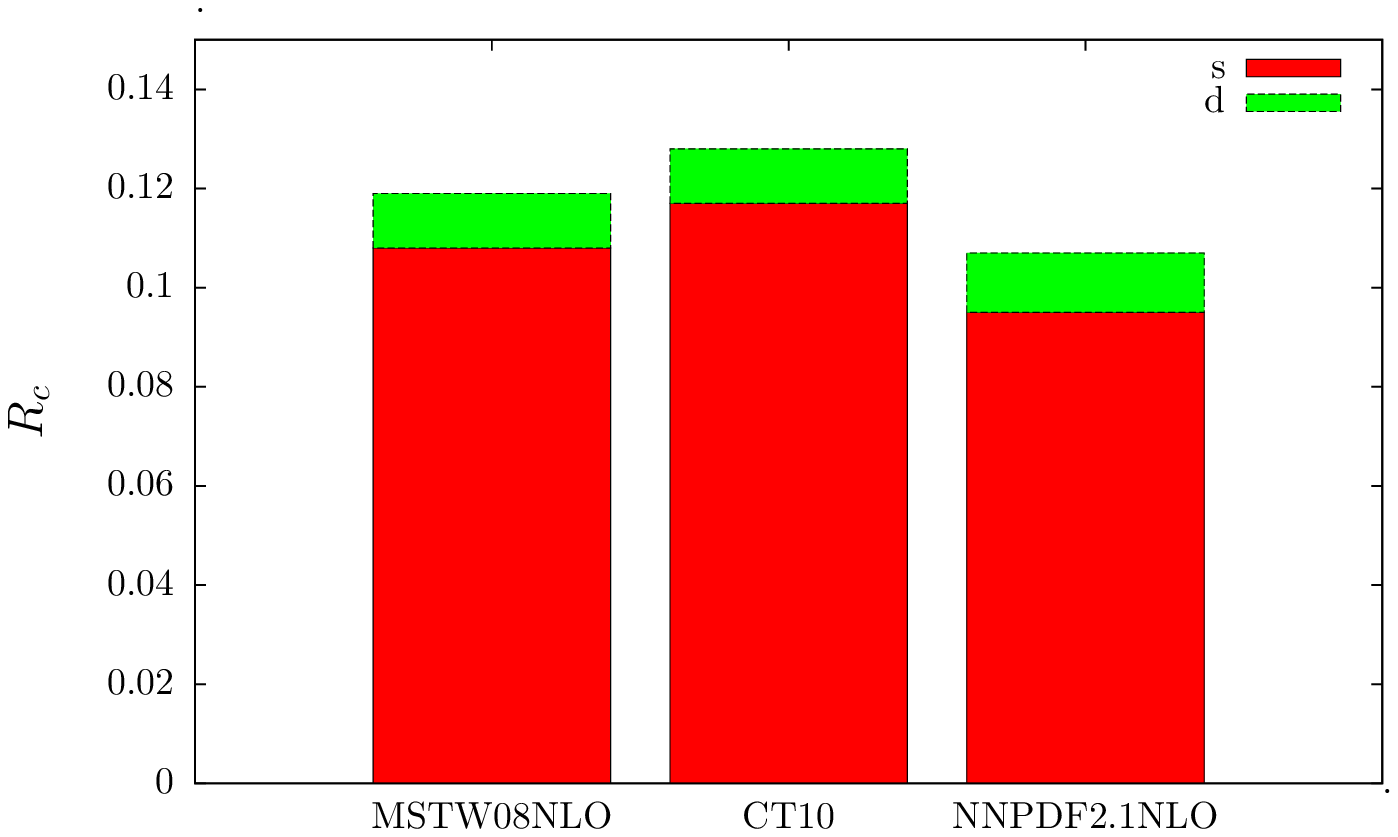}
\caption{Effect of $s$ and $d$ quarks on $R_c$ using NLO PDFs (LO processes).}
\label{NLOpdf2}
\end{minipage}
\end{figure}

The relative contributions of initial-state $s$ and $d$ quarks to $R_c^\pm$ and $R_c$ are illustrated in  Figs.~\ref{NLOpdf1} and \ref{NLOpdf2}.
Note that these results are obtained using LO expressions for the subprocesses, but with NLO PDFs. 
The additional NLO subprocesses, i.e. involving different combinations of initial partons (e.g. $qq,gg$) 
compared to those in Fig.~\ref{Feync}, are of course included in the full NLO calculation, but beyond LO there is 
no unambiguous separation of the cross section into  $s-$ or $d-$quark flavour contributions because of gluon splittings. 
These results are therefore to be used only as a schematic guide in 
determining the relative importance of $s-$ and $d-$quarks in $W+c$ production. 

 For CT10 $s=\bar{s}$ and 
therefore the fact that $R^{\pm}_c<1$ is due entirely to the difference between $d$ 
and $\bar{d}$. NNPDF2.1 does have an asymmetric strange sea, $s-\bar{s} \neq 0$,
 but the asymmetry is very small in the $x, Q^2$ region of interest for this process and therefore  
$R^{\pm}_c\neq 1$ is again determined mainly by the $d$, $\bar{d}$ asymmetry. Finally, for MSTW2008 
the strange asymmetry is larger contributing significantly to $R^{\pm}_c$. 

\begin{figure}[ht]
\begin{minipage}[b]{0.5\linewidth}
\centering
\includegraphics[scale=0.32,angle=-90]{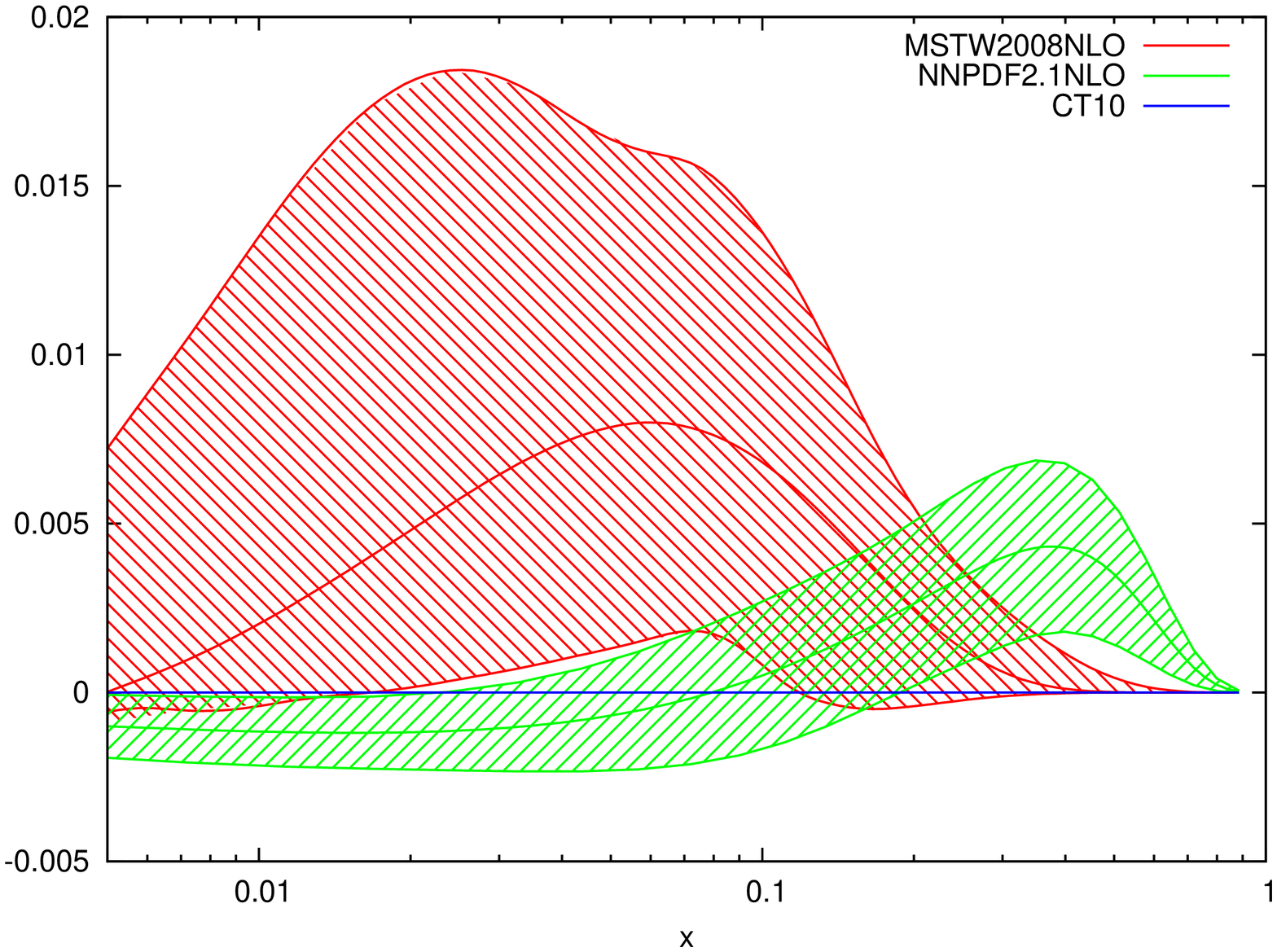} 
\put(-212,-82){\rotatebox{90}{\large{$xs_v$}}}
\caption{Strange valence distribution for NLO PDFs at $Q=M_W$. }
\label{svalMW}
\end{minipage}
\hspace{0.5cm}
\begin{minipage}[b]{0.5\linewidth}
\centering
\includegraphics[scale=0.32,angle=-90]{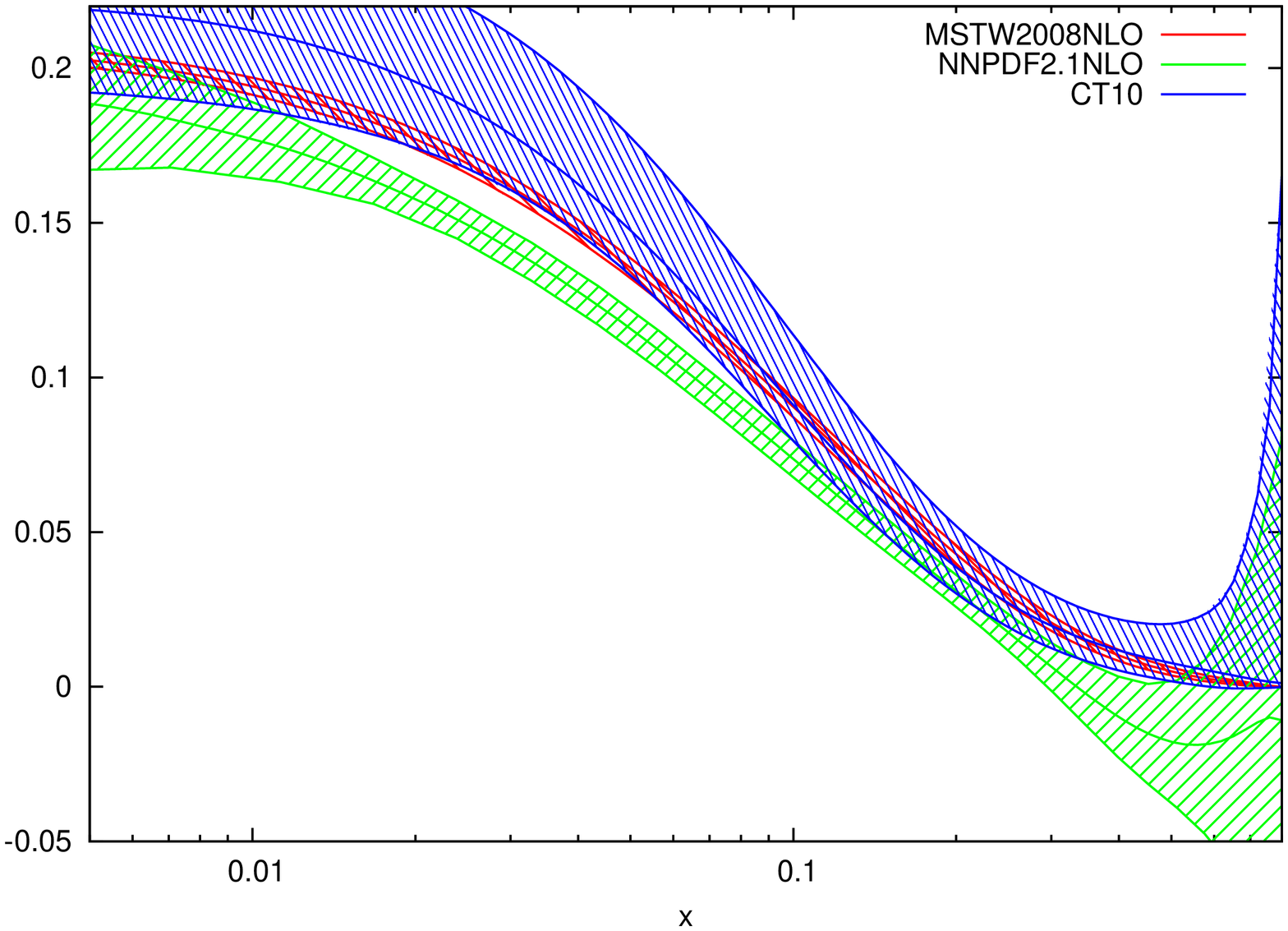}
\put(-220,-92){\rotatebox{90}{\large{$\frac{s+\bar{s}}{\Sigma(q+\bar{q})}$}}}
\caption{NLO PDF ratio of $s+\bar{s}$ to $\Sigma(q+\bar{q})$ at $Q=M_W$.}
\label{strerr}
\end{minipage}
\end{figure}

The strange asymmetry $s-\bar s$ for $Q=M_W$, the relevant scale for this process, 
is shown in Fig.~\ref{svalMW}, including the 68\%cl uncertainty band in the case of MSTW2008NLO and NNPDF2.1. 
The strange asymmetry in both of these sets is constrained by the CCFR and NuTeV dimuon $\nu N$ and $\bar\nu N$ DIS 
data~\cite{Goncharov:2001qe} 
in the global fit. These data slightly prefer an asymmetric strange sea in the $x$ range $0.03 - 0.3$, although the 
CT10 symmetric choice of $s=\bar s$ is also consistent with the data within errors. In the MSTW2008NLO fit, 
the choice of parametrisation drives the 
relatively large positive asymmetry in the range $x \sim 0.01 - 0.1$. There is no such strong parameterisation 
dependence in the NNPDF2.1 fit. 
A precise measurement of the ratio $R_c^\pm$, combined with an improved knowledge of the $d, \bar d$ difference, could therefore provide important new 
information on $s_V$ at small $x$.

The ratio $R_c$ can be used as a measure of the {\em total} strangeness of the proton, 
and to the extent that these $W+$jet cross sections are dominated by $qg$ scattering we can expect
 $R_c\sim\frac{s+\bar{s}}{\Sigma(q+\bar{q})}$. 
 For our three sets of NLO PDFs this ratio at scale $Q=M_W$ is shown in Fig.~\ref{strerr}. 
The ordering of the $R_c$ values and the  relative size of the PDF uncertainties  for the different PDF sets agrees qualitatively with 
the corresponding values of the quark ratio at $x\sim 0.06$, the average value of the incoming quark $x$ for this 
collider energy and choice of cuts. 
The  MSTW2008NLO strange-quark error band is much narrower than that of the other sets because of 
the implicit assumption in the MSTW global fit that all sea quarks have the same 
universal $q_i(x,Q_0^2) \sim x^\delta$ behaviour as $x \to 0$, with $\delta$ determined quite 
precisely by the fit to the HERA small-$x$ structure function data.
 
\begin{figure}[h]
\begin{minipage}[b]{0.5\linewidth}
 \centering
\includegraphics[trim=1cm 0 0 0, scale=0.56]{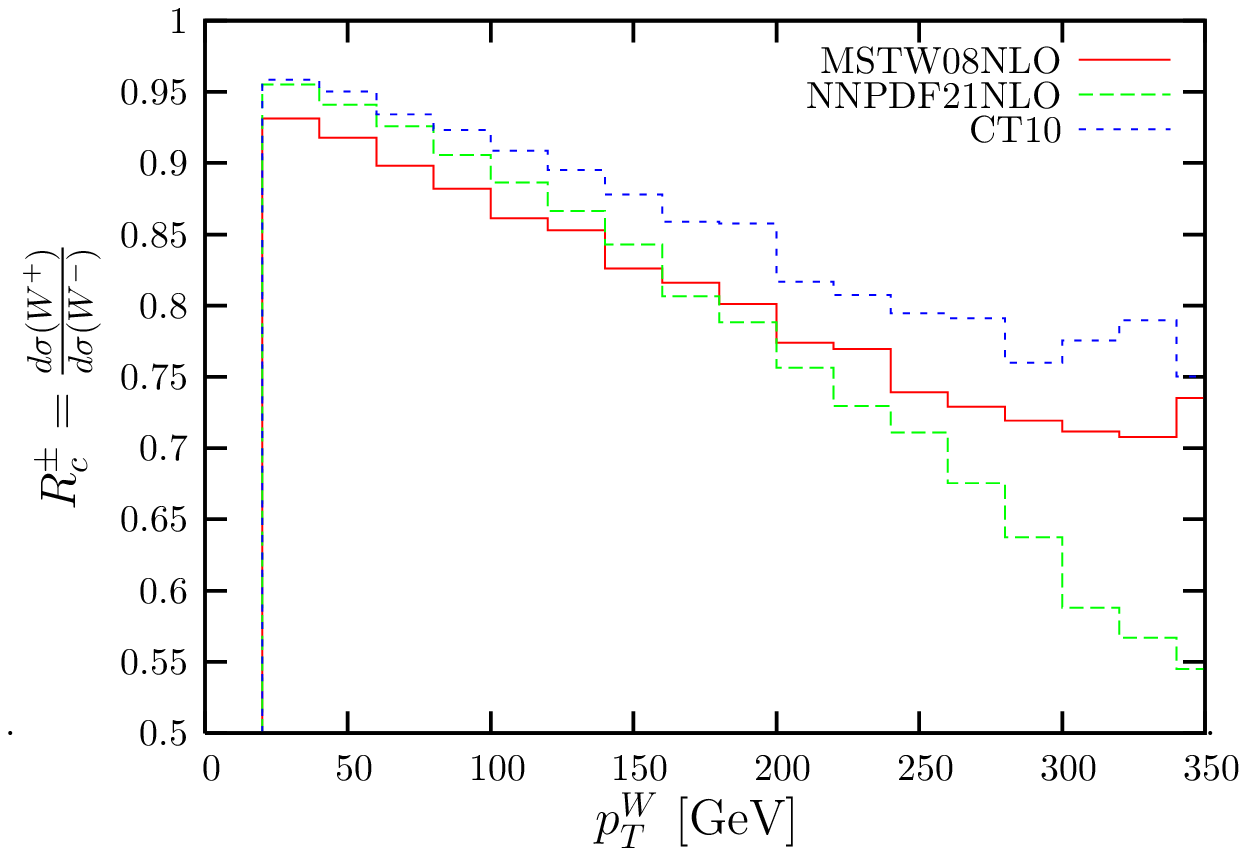}
\caption{Dependence of $R^{\pm}_c$ on $p_T^W$ using NLO PDFs.}
\label{ratioc}
\end{minipage}
\hspace{0.5cm}
\begin{minipage}[b]{0.5\linewidth}
\centering
\includegraphics[trim=1cm 0 0 0,scale=0.56]{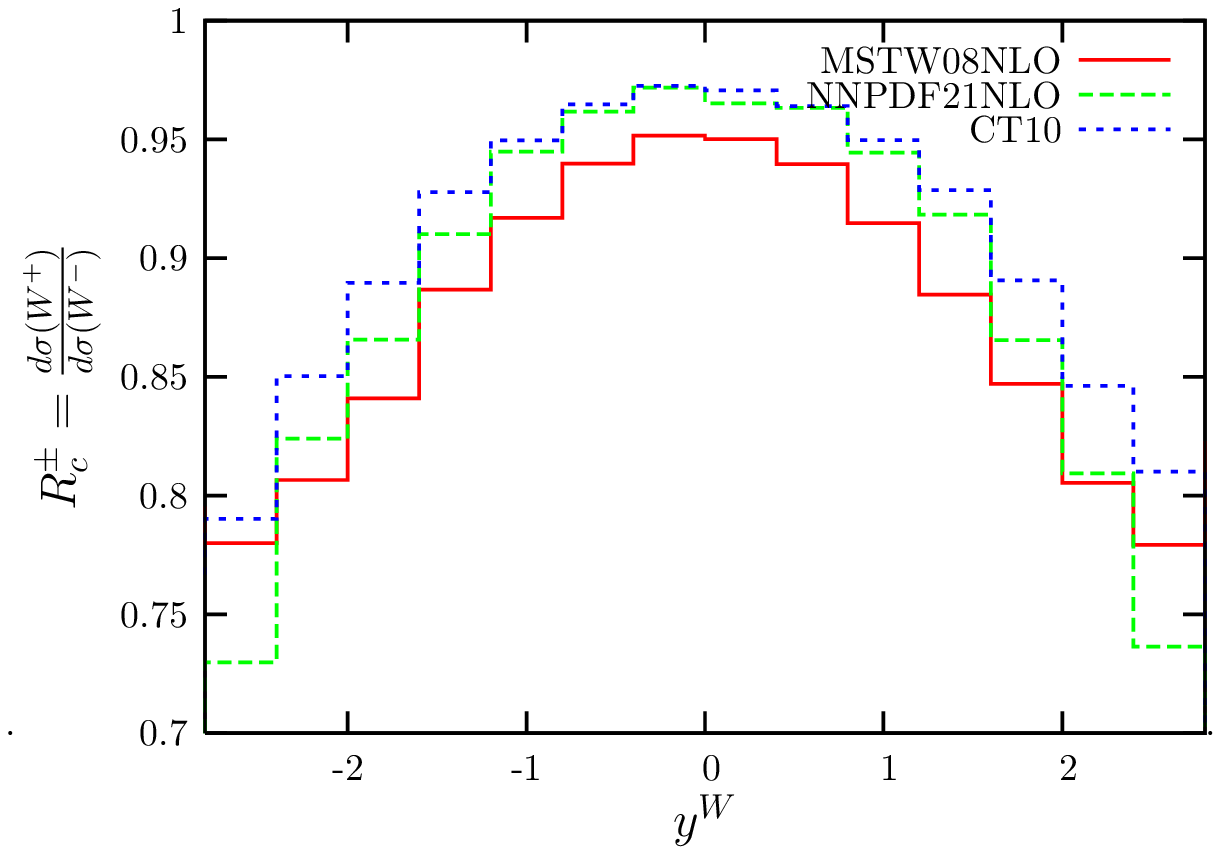}
\caption{Dependence of $R^{\pm}_c$ on $y^W$ using NLO PDFs.}
\label{Wrap}
\end{minipage}
\end{figure}

The ratios $R^{\pm}_c$ and $R_c$ can also be considered 
as distributions of kinematic observables, e.g. the $W$ transverse momentum as shown at LO 
(using NLO PDFs) in Fig.~\ref{ratioc} for $R^{\pm}_c$. In contrast to $R^{\pm}$, which is related 
to the $u/d$ ratio at high $x$ and therefore {\em increases} with $p_T^W$, $R^{\pm}_c$ is a {\em decreasing} 
function  of $p_T^W$ driven by the dominance of the valence $d-$quark at high $x$ over the other parton 
distributions involved, see Eq.~(\ref{eq:sd}). The rapid 
drop for NNPDF is a result of both $\bar{d}/d$ at large $x$ and also the increasing value of $s_V$ 
at large $x$ as shown in Fig.~\ref{svalMW}. The large differences between the predictions of the various PDF 
sets in the region of high $p_T^W$ clearly illustrate
 the potential of using $R^{\pm}_c$ as a PDF discriminator. Similar conclusions can be drawn by considering 
$R^{\pm}_c$ as a function of the $W$ rapidity as shown in Fig.~\ref{Wrap}. The differential distribution of the lepton rapidity is measured by CMS in \cite{CMS:qwa}.

\section{Predictions for $\mathbf{\sigma(Z+c)}$}
Even though the corresponding cross sections with the $W$ boson replaced by a $Z$ boson are significantly smaller, 
especially when account is taken of the difference in the leptonic branching ratios,  with a sufficiently large data sample 
a similar analysis can be performed. 

We first consider the ratio
\begin{equation}
R_c^{Z}=\frac{\sigma(Z+c)}{\sigma(Z+\textrm{jet})} ,
\end{equation}
where the $c$ in the numerator here refers to either a $c$ or a $\bar c$ jet. Defining a similar set of experimental cuts:
 $p_T^{\rm{jet}}>20$ GeV, $|\eta^{\rm{jet}}|<2.1$, $p_T^{\rm{lepton}}>25$ GeV, $|\eta^{\rm{lepton}}|<2.1$, $R^{jj}=0.5$, $R^{lj}=0.3$ and $60<m_{ll}<120$ GeV (to suppress the photon contribution), 
gives the NLO ratio predictions  shown in Table~\ref{Zres}, now with the QCD scales set to $M_Z$. 

\begin{table}[t]
\begin{minipage}[t]{.5\textwidth }
\vspace{0pt}
\begin{center}
\renewcommand*{\arraystretch}{1.4}
    \begin{tabular}{ | c | c |}
    \hline
  PDF set  &  $R_c^Z$ \\ \hline
   CT10  &   0.0619$^{+0.0032}_{-0.0032}$  \\ \hline
MSTW2008NLO & $0.0640^{+0.0014}_{-0.0016}$  \\ \hline
NNPDF2.1NLO & 0.0660$\pm$0.0013 \\ \hline
GJR08 & $0.0611\pm$0.0011\\ \hline
ABKM09 & 0.0605$\pm$0.0019  \\ \hline
   \end{tabular}
\end{center}
\captionof{table}{Comparison of $R_c^Z$ NLO predictions for the different PDF sets, with 68\%cl PDF uncertainties.}
\label{Zres}
\end{minipage}%
\hspace{1cm}
\begin{minipage}[t]{.5\textwidth}
\vspace{0pt}
\centering
\includegraphics[trim=0 1cm 0 0,scale=0.3,angle=-90]{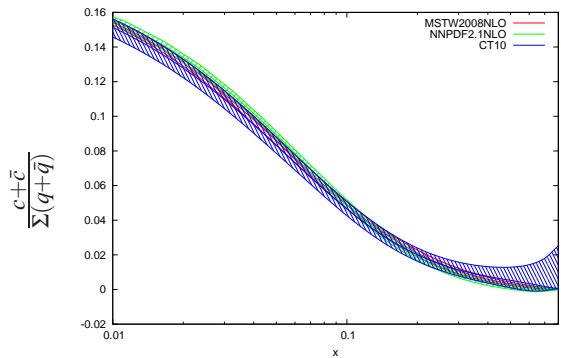} 
\put(-215,-88){\rotatebox{90}{\large{$\frac{c+\bar{c}}{\Sigma(q+\bar{q})}$}}}
\captionof{figure}{Charm quark fraction $(c+\bar{c}) / \Sigma(q+\bar{q})$ at $Q=M_Z$ for NLO PDFs.}
\label{charm1}
\end{minipage}
\end{table}

In principle $R_c^Z$ provides direct information on the charm content of the proton, complementary to that obtained from 
DIS experiments via $F_2^c$ \cite{Martin:2009iq}.  
We note that the differences between the predictions of different PDF sets are much smaller than for the strange quark distributions, presumably because 
in all these global fits the charm distributions arise perturbatively from $g \to c \bar c$ splitting, with the small-$x$ gluon well 
determined from the HERA structure function data. This can be seen in Fig.~\ref{charm1}, 
which compares the ratio of charm quarks to all quarks for the three PDF sets. By analogy with $R_c$, we expect $R_c^Z\sim \frac{c+\bar{c}}{\sum (q+\bar{q})}$. With PDF errors taken into account,
the use of $R_c^Z$ as a PDF discriminator will require a very precise measurement. 

We can also consider the (charm) charge asymmetry ratio: 
\begin{equation}
R_c^{\pm}(Z)=\frac{\sigma(Z+\bar{c})}{\sigma(Z+c)}.
\label{eq:Zcratio}
\end{equation}
$R_c^{\pm}(Z)$ is automatically equal to 1 if $c = \bar c$ 
in the initial state, which is the case for all the PDF sets considered here. However this symmetry does not necessarily hold
if we allow for an {\em intrinsic} charm component~\cite{Brodsky:1980pb}. 
PDF studies incorporating intrinsic charm, see for example~\cite{Pumplin:2007wg,Martin:2009iq}, suggest that it is probably a small effect 
compared to perturbatively generated charm, particularly at the small $x$ values relevant to the LHC. Recently the prospects of searching for intrinsic charm at the LHC using the prompt photon plus charm cross section have been presented in \cite{Bednyakov:2013zta}.

We can also study the ratios:\begin{equation}
R_c^{WZ}=\frac{\sigma(Z+c)}{\sigma(W+c)}\,\,\,\, \textrm{and}\,\,\,\,\, R^{WZ}=\frac{\sigma(Z+\textrm{jet})}{\sigma(W+\textrm{jet})},
\end{equation}
with $R^{WZ}$ measured by ATLAS in \cite{Aad:2011xn}.
The NLO predictions for MSTW2008NLO are 0.045 and 0.082 respectively, 
for the selection cuts described above for $W$ and $Z$ including leptonic decays.

\section{Conclusions}
We have investigated charm production in association with $W$ and $Z$ bosons at the LHC. 
and showed predictions relevant to the recent CMS analysis for $W$ bosons. 
Precise measurements of the ratios $R_c$ and $R_c^{\pm}$ 
 can provide useful information on the strange content of the proton, and in particular the asymmetry between $s$ and $\bar{s}$ at small $x$ and high $Q^2$.
We have also shown results for differential distributions that provide additional information on the $x$ dependence of the strange and anti-strange quark distributions.
We also propose a measurement of the corresponding ratio for Z bosons, $R_c^Z$, which can be used as a measure of the 
charm content of the proton.

\end{document}